\documentclass[twocolumn,aps,prl,showpacs]{revtex4-1}
\usepackage{color}
\usepackage{amsmath}
\usepackage{amssymb}
\usepackage{newlfont}
\usepackage{amsfonts}

\newtheorem{theo}{Theorem}

\newcommand{\tr}{\mbox{Tr}}
\newcommand{\bra}[1]{\mbox{$\langle #1 |$}}
\newcommand{\ket}[1]{\mbox{$| #1 \rangle$}}
\newcommand{\bk}[2]{\ensuremath{\langle #1 | #2 \rangle}}

\begin{document}
\title{Critical sets of the total variance of state detect all SLOCC entanglement
classes}

\author{Adam Sawicki$^{1,2}$, Micha{\l} Oszmaniec$^2$, and Marek Ku\'s$^2$}

\address{$^1$School of Mathematics, University of Bristol,
University Walk, Bristol BS8 1TW, UK, \\
$^2$Center for Theoretical Physics PAS, Al.\ Lotnik\'ow 32/46, 02-668
Warszawa, Poland}


\date{}
\begin{abstract}
We present a general algorithm for finding all classes of pure multiparticle
states equivalent under Stochastic Local Operations and Classsical
Communication (SLOCC).  We parametrize all SLOCC classes by the critical sets
of the total variance function. Our method works for arbitrary systems of
distinguishable and indistinguishable particles. We also show how to calculate
the Morse indices of critical points which have the interpretation of the
number of independent non-local perturbations increasing the variance and hence
entanglement of a state. We illustrate our method by two examples.
\end{abstract}
\maketitle

The problem of classifying pure states up to Stochastic Local Operations
assisted by Classical Communication (SLOCC) has been intensely studied during
the last decade by many authors
\cite{Dur00,kunraty02,chinczyki1,operational}. Although the SLOCC classes are
known for some particular systems, for example in the cases of three or four
qubits, the general method allowing similar derivations for arbitrary system
of many particles which treats in the unified way distinguishable and
indistinguishable particles has been missing. In this letter we provide such
an algorithm. It is based on studying the structure of critical points of the
total variance of a state with respect to local unitary (LU) operations.
Since the total variance of the state can be interpreted as an entanglement
measure \cite{klyachko08} our approach has a clear physical meaning. Although
our results are based on some relatively advanced mathematical tools, the
final algorithm we propose reduces the problem to simple computations
involving diagonalized reduced density matrices.

The paper is organized as follows. First we give a definition of the total
variance of a pure state, $\mathrm{Var}([\psi])$. Next we discuss a
connection between the concept of the momentum map and the reduced
one-particle density matrices. Following this we show that the total variance
of a state is, up to some unimportant additive constant, the norm squared of
the momentum map. This observation allows us to identify the critical points
of $\mathrm{Var}([\psi])$, i.e.\ points $[\psi]$ in $\mathbb{P}(\mathcal{H})$
for which $d\mathrm{Var}([\psi])=0$, with the critical points of the momentum
map. The later turn out to be well understood in Geometric Invariant Theory
(GIT) \cite{Ness84,Kirwan82}. Using its results we show that for each SLOCC
class of pure states the restriction of $\mathrm{Var}([\psi])$ to it attains
maximum on exactly one orbit of the local unitary action. This orbit contains
the most entangled representatives of a given SLOCC class. Finally, we
describe how to calculate the Morse index of $\mathrm{Var}([\psi])$ at
critical points and explain that it has a meaningful interpretation as the
number of independent non-local perturbations which increase the total
variance, $\mathrm{Var}([\psi])$ thus, consequently, also its entanglement.
The whole idea is illustrated by familiar examples.

Throughout the paper we consider the system of $L$ identical particles which
can be either distinguishable or indistinguishable. For the former the
Hilbert space is the tensor product of $L$ copies of
$\mathbb{C}^{N}$,~$\mathcal{H}_{d}=\mathbb{C}^{N}\otimes\ldots\otimes\mathbb{C}^{N}$.
Local unitary operations are represented by the direct product of $L$ copies
of $SU(N)$, $K=SU(N)^{\times L},$ and SLOCC operations by of $L$ copies of
$SL(N)$,~$G=SL(N)^{\times L}$, (see
\cite{Dur00,kunraty02,chinczyki1,operational}). The action of an element
$(A_{1},\,\ldots,\, A_{L})$ of either $K$ or $G$ on the vector
$\ket{\psi}\in\mathcal{H}_{d}$ is given by
\begin{gather}
\ket{\psi}\mapsto A_{1}\otimes\ldots\otimes A_{L}\ket{\psi}\ .\label{eq:dist}
\end{gather}
For indistinguishable particles, i.e.\ for systems of $L$ bosons or $L$
fermions the relevant Hilbert spaces are fully symmetric or fully antisymmetric
parts of the full tensor product, i.e.
$\mathcal{H}_{b}=\mathrm{Sym^{L}}\left(\mathbb{C}^{N}\right)$ and
$\mathcal{H}_{f}=\bigwedge^{L}\left(\mathbb{C}^{N}\right)$. The local groups
corresponding to LU and SLOCC operations are $K=SU(N)$ and $G=SL(N)$ and the
action of an element $A\in K$ (or $A\in G$) on $\ket{\psi}\in\mathcal{H}_{b/f}$
is given by (\ref{eq:dist}) with $A_{k}=A$. Because the global phase factor and
the normalization are physically irrelevant it is useful to identify pure
states with elements of the complex projective space
$\mathbb{P}\left(\mathcal{H}\right)$. To this end we identify vectors from
$\mathcal{H}$ that differ by a nonzero complex scalar factor and denote by
$\left[\psi\right]\in\mathbb{P}\left(\mathcal{H}\right)$ the state
corresponding to vectors $\ket{\psi}\in\mathcal{H}$ under this identification.
Because groups $K$ and $G$ act linearly on $\mathcal{H}$ they also act
naturally on $\mathbb{P}\left(\mathcal{H}\right)$. Two pure states
$\left[\psi\right],\,\left[\phi\right]\in\mathcal{\mathbb{P}\left(\mathcal{H}\right)}$
are SLOCC equivalent if and only if $\ket\psi$ and $\ket\phi$ can be connected
by the action of $G$ on the level of $\mathcal{H}$, i.e.
$g.\ket{\psi}=e^{i\alpha}\ket{\phi}$, $\alpha\in[0,\,2\pi)$, for some $g\in G$.
Actions of $K$ and $G$ on $\mathcal{H}$ induce actions of the corresponding Lie
algebras.
The action of $\alpha=\left(\alpha_{1},\ldots,\alpha_{L}\right)$ in either
$\mathfrak{su}(N)^{\oplus L}$ or $\mathfrak{sl}(N)^{\oplus L}$ on the vector
$\ket\psi\in\mathcal{H}$ is given by
$\alpha.\ket{\psi}=\left(\alpha_{1}\otimes I_{N}\otimes\ldots\otimes I_{N}.
+\ldots +I_{N}\otimes\ldots\otimes I_{N}\otimes\alpha_{L}\right)\ket{\psi}.
$
For indistinguishable particles $\alpha_k=\alpha\in\mathfrak{su}(N)$ (or
$\alpha\in\mathfrak{sl}(N)$).

After fixing notation we can define the total variance of the state
$\left[\psi\right]\in\mathbb{P}\mathcal{\left(H\right)}$, first introduced in
\cite{klyachko08}
\begin{gather}
\mathrm{Var}([\psi])=\frac{1}{\bk{\psi}{\psi}}\sum_{i=1}^{\dim\,
K}\left(\bra{\psi}X_{i}{}^{2}\ket{\psi}+
\frac{1}{\bk{\psi}{\psi}}\bra{\psi}X_{i}\ket{\psi}^{2}\right),
\label{eq:variance}
\end{gather}
where $\{X_{i}\}_{i=1}^{\dim\, K}$ is an orthonormal (with respect to the
Hilbert-Schmidt scalar product $(A,B)=\tr(A^\dagger B)$) basis of
$\mathfrak{k}$ represented as matrices acting in $\mathcal{H}$\footnote{We
define Lie algebra $\mathfrak{k}$ of a Lie group $K$ by the physics convention:
$X\in\mathfrak{k}$ if and only if $\mathrm{exp}\left(iX\right)\in K$. Therefore
Lie algebra $\mathfrak{su}(N)$ consists of traceless hermitian
operators. %
}. The expression (\ref{eq:variance}) is simply the sum of variances of
'local' observables $X_{i}$ calculated in the state $\left[\psi\right]$. It
can be checked that $\mathrm{Var}([\psi])$ is $K$ - invariant and attains its
minimum precisely on the set of separable states. In fact, it can be used as
a measure of entanglement \cite{klyachko08}.

In order to proceed we briefly describe the concept of the momentum map. This
is a map which encodes information about all first integrals of a given
classical Hamiltonian system with symmetries. For example, for a Hamiltonian
system with the rotational symmetry, i.e.\ when the symmetry group is
$K=SO(3)$, the first integrals are three components of the angular momentum
corresponding to the invariance of the system with respect to infinitesimal
rotations along axes $x$, $y$, $z$. The infinitesimal rotations generate the
Lie algebra $\mathfrak{so}(3)$ of $SO(3)$. There are many possible choices of
basis for $\mathfrak{so}(3)$, each corresponds to different choice of rotation
axes and each gives three first integrals. The mathematical object which
encodes information about all first integrals for all possible choices of
generators of $\mathfrak{s0}(3)$ is a map
$\mu:M\rightarrow\mathfrak{so}(3)^{\ast}$, i.e.\ a map from the phase space $M$
to the space of linear functionals on the Lie algebra $\mathfrak{so}(3)$ of the
group $SO(3)$. For every infinitesimal symmetry $\xi\in\mathfrak{so}(3)$ one
can find the corresponding first integral by evaluating
$\mu_{\xi}(x)=\mu(x)(\xi)$. This idea can be generalized to arbitrary $K$ and
the corresponding map $\mu:M\rightarrow\mathfrak{k}^{\ast}$ is called momentum
map \cite{GS90}.

It is well known that $\mathbb{P}\left(\mathcal{H}\right)$
posses a natural phase space structure%
 \footnote{By the term ``phase space structure'' of $\mathcal{M}$ we actually
mean its symplectic structure.%
}. Moreover, for considered composite systems we can treat LU operations as
symmetries of the system as they do not change entanglement. Remarkably, the
momentum map for both distinguishable and indistinguishable particles has a
clear quantum mechanical meaning. It is a map which assigns to each state its
reduced one-particle density matrices. For distinguishable particles we have
\begin{gather}
\mu(\left[\psi\right])=\left(\rho_{1}([\psi])-
\frac{1}{N}I_{N},\ldots,\,\rho_{N}([\psi])-\frac{1}{N}I_{N}\right)\,,
\label{eq:mudist}
\end{gather}
 where $\rho_{i}([\psi])$ is the normalized $i$-th reduced one-particle
density matrix of the vector state $\ket{\psi}$. For indistinguishable
particles we have:
\begin{gather}
\mu(\left[\psi\right])=\rho_{1}([\psi])-\frac{1}{N}I_{N}\,.\label{eq:mubf}
\end{gather}
 Since $\mu(\left[\psi\right])$ is an element of $\mathfrak{k}$
we can calculate its Hilbert-Schmidt norm. It is easy to see that:
\begin{gather}
\left\Vert \mu(\left[\psi\right])\right\Vert
^{2}=\frac{1}{{\bk{\psi}{\psi}}^{2}}\sum_{i=1}^{\dim\,
K}{\bra{\psi}X_{i}\ket{\psi}}^{2}\label{eq:norm sq}
\end{gather}
We also notice that $\mathcal{C}_{2}=\sum_{i=1}^{\dim\, K}X_{i}^{2}$ is the
representation of the second order Casimir operator \cite{BR80}, a
generalization to other Lie algebras of the squared total angular momentum
relevant for the above mentioned example of rotationally invariant system.
The operator $C_2$ commutes with each $X_{i}$. Since the group $K$ acts on
$\mathcal{H}$ irreducibly, it is proportional to the identity operator with
the proportionality constant
$c=\frac{\bra{\psi}\mathcal{C}_{2}\ket{\psi}}{\bk{\psi}{\psi}}$ . Making use
of this formula \eqref{eq:norm sq} we get the final expression for the total
variance of a state:
\begin{gather}
\mathrm{Var}(\left[\psi\right])=c-\left\Vert
\mu(\left[\psi\right])\right\Vert ^{2}\,.\label{final}
\end{gather}
Notice that by (\ref{final}) the critical points $\mathrm{Var}([\psi])$ are
exactly the critical points of $||\mu([\psi])||^{2}$. The later where intensely
studied in GIT context in the 80'. Specifically, the following theorem is the
reformulation of the general result \cite{Kirwan82,Ness84}

\begin{theo}\label{th1} Assume that $[\psi]$ is a critical point
of $\mathrm{Var}:\mathbb{P}(\mathcal{H})\rightarrow\mathbb{R}$. Then the
restriction of the total variance function $\mathrm{Var}$ to the SLOCC class
of $[\psi]$, i.e.\ $G$-orbit through $[\psi]$, attains its maximum value on a
unique $K$-orbit, which is the orbit $K.[\psi]$ through the state $[\psi]$.
\end{theo}
In general it may happen that a particular $G$-orbit does not contain any
critical point of $\mathrm{Var}:\mathbb{P}(\mathcal{H})\rightarrow\mathbb{R}$.
Nevertheless, as we show in \cite{sawicki11a}, such an orbit always contains in
its closure the $K$-orbit which is critical for $\mathrm{Var}([\psi])$. The
mathematical details of this construction are rather subtle and we will not
discuss them here. We would like to emphasize, however, that knowing critical $K$-orbits
allows the reconstruction of these $G$-orbits which do not contain any
critical point of $\mathrm{Var}$ (see \cite{sawicki11a}). Under the above assumptions
Theorem~\ref{th1} can be interpreted as the statement saying that the critical
sets of the total variance function of a many-particle system parametrize all
SLOCC classes of states. Moreover the distinguished local unitary orbit on
which $\mathrm{Var}$ attains a maximum contains maximally entangled
representatives of a given SLOCC class. Although we already know that the total
variance cannot increase when we perturb the state from this orbit with SLOCC
operations, it is still not clear what happens when the perturbation is
non-local. This information is stored in the Morse index of a critical point
i.e.\ the number of negative eigenvalues of the Hessian of $\left\Vert
\mu\left(\left[\psi\right]\right)\right\Vert ^{2}$ calculated at $[\psi]$. We
will investigate this idea on two examples. First, however we need some
effective way for finding critical points of $||\mu(\left[\psi\right])||^{2}$
and calculating the Morse index. In \cite{sawicki11a} we prove the following
theorem
\begin{theo}\label{thm2}
\begin{enumerate}
\item The state $\left[\psi\right]$ is a critical point of
    $\mathrm{Var}:\mathbb{P}(\mathcal{H})\rightarrow\mathbb{R}$ if and
    only if the corresponding vector $\ket{\psi}$ is an eigenvector of
    $\mu(\left[\psi\right])$, i.e.\ for some $\lambda\in\mathbb{R}$ we
    have
\begin{gather}
\mu(\left[\psi\right])\ket{\psi}=\lambda\ket{\psi}\label{eq:eigen}.
\end{gather}
\item The Morse index of $\left\Vert \mu\left(\left[\psi\right]\right)\right\Vert ^{2}$
at a critical point $[\psi]$ is the the Morse index of the function
$f:\mathcal{H}\rightarrow\mathbb{R}$,
\begin{gather}
f(\ket{\phi})=\frac{\bra{\phi}\mu(\left[\psi\right])\ket{\phi}}{\bk{\phi}{\phi}},\label{eq:hess}
\end{gather}
calculated at $\ket{\phi}=\ket{\psi}$ and restricted to the orthogonal
complement (with respect to the usual inner product on $\mathcal{H}$) of
the subspace $\mathfrak{g}.\ket{\psi}$
\end{enumerate}
\end{theo} In Equations \eqref{eq:eigen} and \eqref{eq:hess} $\mu(\left[\psi\right])$
is understood as an operator acting on $\mathcal{H}$. Expression
$\mathfrak{g}.\ket{\psi}$ denotes the image of the vector $\ket{\psi}$ under
the action of the whole $\mathfrak{g}$ represented on $\mathcal{H}$.

Using (\ref{eq:eigen}) we can divide the problem of finding critical points
of $\mathrm{Var}([\psi])$ into two conceptually different parts. First, we
notice that condition (\ref{eq:eigen}) is automatically satisfied if
$\mu([\psi])=0$. This corresponds to states with maximally mixed reduced
one-particle density matrices. Since the set $\mu^{-1}(0)$ is $K$-invariant,
each $K$-orbit contained in $\mu^{-1}(0)$ represents a different SLOCC class
of states. By splitting $\mu^{-1}(0)$ into disjoint orbits of $K$ and picking
exactly one state from each orbit we get a parametrization of all SLOCC
classes of this type. These classes contain almost all states
\cite{sawicki11a}. For effective calculations it is desirable to have some
canonical form of the state. In $N$-particle case such a form was given in
\cite{VDM03}. The remaining critical points of $\mathrm{Var}(\psi)$ stem from
(\ref{eq:eigen}) when $\mu([\psi])\neq0$. The following two remarks are
crucial in this case (see \cite{sawicki11a} for a detailed discussion)

1. For any $\ket{\psi}$ there exists LU operator
    $U=U_{1}\otimes\ldots\otimes U_{L}$ such that $\mu([\psi^{\prime}])$,
    where $\ket{\psi^{\prime}}=U\ket{\psi}$\textcolor{blue}{,} is given by
    \eqref{eq:mudist} with diagonal reduced density matrices
    $\rho_{i}([\psi^{\prime}])$ whose eigenvalues are arranged in
    non\textcolor{blue}{-}increasing order; the matrices $U_{i}\in SU(N)$
    diagonalize $\rho_{i}([\psi])$. Notice that since critical sets of
    $\mathrm{Var}(\psi)$ are $K$ - invariant it implies that we can
    consider equation (\ref{eq:eigen}) assuming that $\psi$ posses the
    above described property of $\psi^{\prime}$. For bosons and fermions
    the relevant operators $U$ also exist and are given by
    $U=U_{1}\otimes\ldots\otimes U_{1}$, where $U_{1}\in SU(N)$ diagonalize
    $\rho_{1}([\psi])$ from (\ref{eq:mubf}) and make its spectrum non
    increasingly ordered.

2. Let $P_{i}(\psi)$ be a point in $\mathbb{R}^{N}$ whose coordinates
    are given by non\textcolor{blue}{-}increasingly ordered spectrum of
    $\rho_{i}([\psi])-\frac{1}{N}I_{N}$. The set
    $\Psi(\mathbb{P}(\mathcal{H}))
    =\{\left(P_{1}(\psi),\ldots,P_{L}(\psi)\right):\,[\psi]\in\mathbb{P}(\mathcal{H})\}$
    is a convex polytope in $\mathbb{R}^{LN}$, the so-called Kirwan or
    momentum polytope. This fact is a simple corollary \cite{SWK12} from
    the general theorem known as the convexity property of the momentum map
    \cite{Kirwan84}. Moreover, in the variety of settings for
    distinguishable and indistinguishable particles the inequalities
    defining this polytope are explicitly known \cite{Higuchi03,Pauli
    principle}.

Using the above remarks the search for remaining critical points of
$\mathrm{Var}(\psi)$ can be reduced to the following procedure. For each
point $P=(P_{1},\ldots,P_{L})$ in the polytope
$\Psi(\mathbb{P}(\mathcal{H}))\setminus\{0\}$ we construct corresponding
operator
\begin{gather*}
\alpha_{P}=\alpha_{P_{1}}\otimes I_{N}\otimes\ldots\otimes I_{N}
+\ldots+I_{N}\otimes\ldots\otimes I_{N}\otimes\alpha_{P_{L}}\,,
\end{gather*}
where $\alpha_{P_{i}}$ is a diagonal matrix with diagonal elements
given by $P_{i}$. Next we are looking for states $[\psi]$ for which
the condition
\begin{gather}
\alpha_{P}\ket{\psi}=\lambda\ket{\psi}\label{eq:poly-mu}
\end{gather}
is satisfied and, at the same time, $\mu(\left[\psi\right])=\alpha_{P}$. The
problem of finding critical points is thus reduced to a study of eigenstates
of the operator $\alpha_{P}$ which is an $N^{L}\times N^{L}$ diagonal matrix
for $P\in\Psi(\mathbb{P}(\mathcal{H}))$. If the diagonal elements of
$\alpha_{P}$ are nondegenerate the corresponding eigenvectors are separable
states. Hence in order to find nontrivial SLOCC classes we are interested in
the situation when the spectrum of $\alpha_{P}$ is degenerate. Remarkably,
the dimensions of the degenerate eigenspaces are typically relatively small
and hence it is quite easy to verify for which states
$\mu(\left[\psi\right])=\alpha_{P}$. Finally, as we explain in
\cite{sawicki11a} it is always possible to divide critical points found in
this way into finite number of families, where each family contains SLOCC
nonequivalent classes of states characterized by fixed reduced one-particle
density matrices.

In order to demonstrate how to use the above method we provide calculations for
two simple examples (three qubits and three 5-state fermions). For cases when
calculation are more elaborate see \cite{sawicki11a}.

\emph{Three qubits}. For the three-qubit system the Hilbert space is
$\mathcal{H}=\mathbb{C}^{2}\otimes\mathbb{C}^{2}\otimes\mathbb{C}^{2}$, the
SLOCC operations are represented by the group $G=SL(2)^{\times 3}$, and the
local unitary operations by the group $K=SU(2)^{\times 3}$. Using the
cannonical form of a three-qubit state \cite{VDM03},
\begin{equation}
\ket{\psi}=p\ket{011}+q\ket{101}+r\ket{110}+s\ket{111}+z\ket{000},
\label{Acin}
\end{equation}
the reduced one-qubit density matrices can be easily calculated. It is then a
matter of straightforward calculations to see that there are exactly two states
of the form (\ref{Acin}) which have maximally mixed reduced one-particle
density matrices. One of them is the GHZ state,
$\ket{\psi_{GHZ}}=\frac{1}{\sqrt{2}}\left(\ket{000}+\ket{111}\right)$, and the
other one is LU equivalent to it. Effectively, the main SLOCC class is the
$G$-orbit through $GHZ$. The remaining classes can be found by considering
Equation~(\ref{eq:poly-mu}), where $\alpha_P$ is $8\times 8$ diagonal matrix.
The Kirwan polytope is given by three polygonal inequalities \cite{Higuchi03}
for the smallest eigenvalues of the reduced one-qubit density matrices and is a
three-dimensional subset of $\mathbb{R}^6$. By direct calculations we find that
the only eigenspaces contributing new critical points are of dimension $1$, $3$
and $4$. They give separable, biseparable and $W$ SLOCC classes, respectively,
i.e.
\begin{gather*}\label{slocc-3-qubit}
\ket{\psi_{W}}=\frac{1}{\sqrt{3}}\left(\ket{100}+\ket{010}+\ket{001}\right),\\
\ket{\psi_{BS1}}=\frac{1}{\sqrt{2}}\left(\ket{100}+\ket{111}\right),\,\,\psi_{BS2}=\frac{1}{\sqrt{2}}\left(\ket{010}+\ket{111}\right),\\
\ket{\psi_{BS3}}=\frac{1}{\sqrt{2}}\left(\ket{001}+\ket{111}\right),\,\,\psi_{SEP}=\ket{000}.
\end{gather*}
Since the state GHZ is in $\mu^{-1}(0)$ the Morse index at it is equal to
zero \cite{sawicki11a}. This means it is impossible to increase the total
variance $\mathrm{Var}$ by any perturbation of $K.\ket{\psi_{GHZ}}$. The case
of $\ket{\psi_{W}}$ is more interesting, namely for the orthogonal complement
of $\mathfrak{g}.\ket{\psi_1}$ we have
\begin{gather*}
\left(\mathfrak{g}.\ket{\psi_{1}}\right)^{\bot}=\mathrm{Span}\left\{
\ket{111},\, i\ket{111}\right\} .
\end{gather*}
It is easy to check that the Hessian of the function $f(\ket{\psi})$  defined
by Eq.~(\ref{eq:hess}) is negative on
$\left(\mathfrak{g}.\ket{\psi_{1}}\right)^{\bot}$ and hence the index equals
2. Similar calculations show that for biseparable states, i.e.\
$\ket{\psi_{BS1}}$, $\ket{\psi_{BS2}}$ and $\ket{\psi_{BS3}}$ we have
$\mathrm{index}(\ket{\psi_{BSk}})=6$ and for separable state
$\mathrm{index}(\ket{\psi_{SEP}})=8$.

\emph{Three five-state fermions}. As a second example let us consider a
system consisting of three fermions with five-dimensional single particle
Hilbert spaces, $\mathcal{H}=\bigwedge^{3}\left(\mathbb{C}^{5}\right)$,
$K=SU\left(5\right)$, $G=SL(5,\,\mathbb{C})$. In $\mathbb{C}^{5}$ we fix an
orthonormal basis $\left\{ \ket 1,\,\ket 2,\,\ket 3,\,\ket 4,\ket 5\right\} $
and chose the (also orthonormal) basis of
$\bigwedge^{3}\left(\mathbb{C}^{5}\right)$:
\begin{equation}
\ket{i_1,i_2,i_3}:=\ket{i_{1}}\wedge\ket{i_{2}}\wedge\ket{i_{3}} \,,1\leq i_{1}<i_{2}<i_{3}\leq5 \,.
\end{equation}
Any $\ket{\psi}\in\mathcal{H}$ has a decomposition:
\begin{equation}
\ket{\psi}=\sum_{1\leq i_{1}<i_{2}<i_{3}\leq5} c_{i_1,i_2,i_3}
\ket{i_1,i_2,i_3}
\end{equation}
where the scalar coefficients fulfill the normalization condition. A system
with $\mathcal{H}=\bigwedge^{3}\left(\mathbb{C}^{5}\right)$ can be considered
as consisting of two holes and the structure of the spectrum of the rescaled
density matrix $\rho\left([\psi]\right)$ is well known \cite{Pauli
principle}. First condition says that the eigenvalues of $\rho$ are bounded
from above by $\frac{1}{3}$. The second one is that the maximal eigenvalue
$\lambda_{max}$ equals exactly $\frac{1}{3}$. The last condition states that
all eigenvalues except for $\lambda_{max}$ are, at least, doubly degenerate.
Using these properties it can be easily shown that there are only two
nonequivalent critical sets of $\mathrm{Var}$. They are parametrized by the
states:
\begin{equation}
\ket{\psi_{1}}=\ket{1,2,3}, \quad
\ket{\psi_{2}}=\frac{1}{\sqrt{2}}\left(\ket{1,2,3}+\ket{1,4,5}\right)\,.
\end{equation}
Using Theorem 2 we find, in the same manner as in the case of three qubits,
that $\mathrm{index}\left(\ket{\psi_{1}}\right)=6$ and
$\mathrm{index}\left(\ket{\psi_{2}}\right)=0$. The latter fact is a consequence
of the observation that $\left(\mathfrak{g}.\ket{\psi_{2}}\right)^{\perp}=0$.
It follows that $G.[\psi_{2}]$ is of the maximal dimension and that the total
variance attains the global maximum on the $K$ orbit passing through
$[\psi_2]$. It is interesting to note that in the case considered $\mu^{-1}(0)$
is empty, that is there are no states in $\mathbb{P}(\mathcal{H})$ that have
maximally mixed reduced density matrix.

Summarizing, we provided an algorithm for finding and classifying SLOCC
equivalent pure states. The algorithm is effective and universally applicable
in various setting including distinguishable and indistinguishable systems of
particles with arbitrary number of components.

We gratefully acknowledge the support of SFB/TR12 Symmetries and Universality
in Mesoscopic Systems program of the Deutsche Forschungsgemeischaft, ERC Grant
QOLAPS, a grant of the Polish National Science Centre under the contract number
DEC-2011/01/M/ST2/00379 and Polish MNiSW grant no. IP2011048471.

After completion of this work, we have learned about independent related work
by Walter, Doran, Gross, and Christandl \cite{walter12}.

\end{document}